\documentclass[conference]{IEEEtran}
\IEEEoverridecommandlockouts

\usepackage[switch]{lineno}
\usepackage[colorlinks = true,
            linkcolor = blue,
            urlcolor  = blue,
            citecolor = blue,
            anchorcolor = blue]{hyperref}

\usepackage{amsmath,amssymb,amsfonts}
\usepackage{algorithmic}
\usepackage{graphicx,subfig,mwe}
\usepackage{tikz}
\usepackage{textcomp}
\usepackage{xcolor}

\usepackage{enumerate}
\usepackage{orcidlink}
\usepackage{wrapfig}
\usepackage{booktabs,tabularx,multirow}
\usepackage[style=ieee]{biblatex}
\def\BibTeX{{\rm B\kern-.05em{\sc i\kern-.025em b}\kern-.08em
    T\kern-.1667em\lower.7ex\hbox{E}\kern-.125emX}}

  \newcommand*\pinkFigureBox[2]{\draw[pink,thick] (#1) rectangle (#2);}
\newenvironment {annotatedFigure}[1]{\centering\begin{tikzpicture}
\node[anchor=south west,inner sep=0] (image) at (0,0) { #1};\begin{scope}[x={(image.south east)},y={(image.north west)}]}{\end{scope}\end{tikzpicture}}

\usepackage{xspace}
\newcommand{\method}{\textit{HeartSpot}\xspace}

\newcommand{\contributions}{
\begin{itemize}
  \item \textbf{Compression} of single images with no learning, giving up to to $11x$ smaller filesize and $32x$ fewer pixels.
  \item \textbf{Speed and Accuracy Gains} up to $+0.01$ area under ROC curve or $9x$ faster training throughput.
  \item \textbf{Privatized} and \textbf{Ante-hoc explainable} compression via a spatial bias prior.  Up to 97\% of pixels are removed while preserved pixels clearly visualize the heart.
  \item \textbf{Post-Hoc Attribution Explanations} without requiring the original image and improved via a quantile filter.
\end{itemize}
}

\begin{document}

\title{\method: Privatized and Explainable Data Compression for Cardiomegaly Detection\\
\thanks{The TAMI project is funded by the European Regional Development Fund (ERDF) through the Programa Operacional Regional do Norte (NORTE 2020) and by National Funds through the Portuguese Foundation for Science and Technology (FCT). I.P. within the scope of the CMU Portugal Program LA/P/0063/2020 and NORTE-01-0247-FEDER-045905.  Alex Gaudio is funded by the FCT, SFRH/BD/143114/2018.
}}

\author{
     \IEEEauthorblockN{
        Elvin Johnson\IEEEauthorrefmark{1},
        Shreshta Mohan\IEEEauthorrefmark{1},
        Alex Gaudio\IEEEauthorrefmark{1}\IEEEauthorrefmark{2}\IEEEauthorrefmark{3}\orcidlink{0000-0003-1380-6620},
        Asim Smailagic\IEEEauthorrefmark{1}\orcidlink{0000-0001-8524-997X},
        Christos Faloutsos\IEEEauthorrefmark{1}\orcidlink{0000-0003-2996-9790},
        Aur\'{e}lio Campilho\IEEEauthorrefmark{2}\IEEEauthorrefmark{3}\orcidlink{0000-0002-5317-6275}
    }
    \IEEEauthorblockA{\IEEEauthorrefmark{1} Carnegie Mellon University, Pittsburgh, PA, USA}
    \IEEEauthorblockA{\IEEEauthorrefmark{2} Faculty of Engineering of the University of Porto, Porto, Portugal}
    \IEEEauthorblockA{\IEEEauthorrefmark{3} INESC TEC, Porto, Portugal}
}

\maketitle

\begin{abstract}
  Advances in data-driven deep learning for chest X-ray image analysis underscore the need for explainability, privacy, large datasets and significant computational resources.
  We frame privacy and explainability as a lossy single-image compression problem to reduce both computational and data requirements without training.
 For Cardiomegaly detection in chest X-ray images, we propose \method and four spatial bias priors.  \method priors define how to sample pixels based on domain knowledge from medical literature and from machines.
\method privatizes chest X-ray images by discarding up to 97\% of pixels, such as those that reveal the shape of the thoracic cage, bones, small lesions and other sensitive features.   \method priors are ante-hoc explainable and give a human-interpretable image of the preserved spatial features that clearly outlines the heart.
\method offers strong compression, with up to $32x$ fewer pixels and $11x$ smaller filesize.  Cardiomegaly detectors using \method are up to $9x$ faster to train or at least as accurate (up to $+.01$ AUC ROC) when compared to a baseline DenseNet121.
 \method is post-hoc explainable by re-using existing attribution methods without requiring access to the original non-privatized image.
  In summary, \method improves speed and accuracy, reduces image size, improves privacy and ensures explainability.
\end{abstract}


\section{Introduction}
Cardiomegaly is a medical condition describing an enlarged heart. It can indicate underlying or life-threatening heart problems.
Medical domain knowledge of Cardiomegaly describes its detection from chest X-ray images as a relation between the volume of the heart to the volume of the thoracic cage.  The cardio-thoracic ratio in a two-dimensional image, for instance, compares the transverse diameter of the heart in pixels to that of the thoracic cage.  A variation compares the area of the heart to the lungs \cite{browne2004extraction}.  Our approach, based on sampling of lines of pixels, encompasses the main idea of both techniques. 

Existing literature on Cardiomegaly detection from chest X-ray considers the problem via classification or segmentation \cite{sogancioglu2020cardiomegaly}.  For classification on the CheXpert \cite{irvin2019chexpert} dataset of chest X-ray images, several standard deep networks were evaluated to offer an expected benchmark \cite{bressem2020comparing}.
The work of \cite{que2018cardioxnet} trains a U-Net to segment the heart and thorax, and then computes the thoracic ratio from image segmentation.   Segmentation improves performance over classification, but requires pixel-wise labels.  The existing literature does not consider privacy or compression.  Our approach improves both efficiency and privacy through single-image compression and is compatible with any of these classification or segmentation approaches.

\begin{figure}[t]
  \centering
  \subfloat[Good Compression\label{fig:1b}]{\includegraphics[width=.45\linewidth]{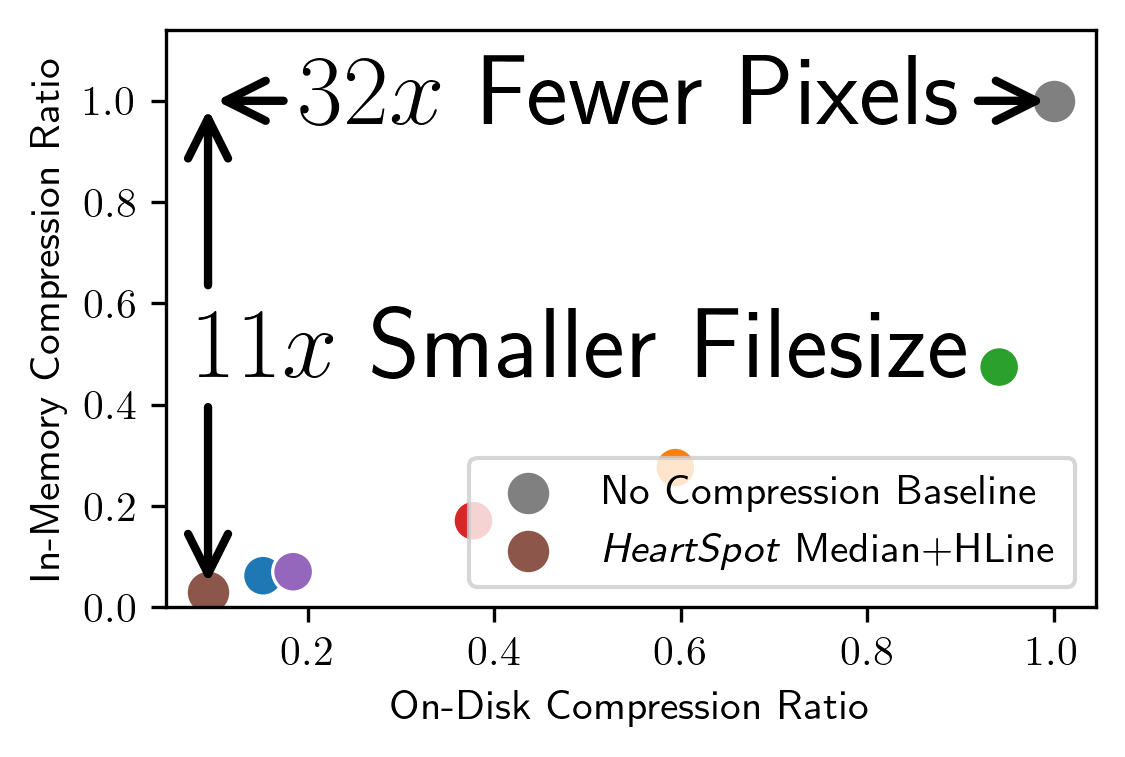}}
  \quad
  \subfloat[Fast and Accurate\label{fig:1a}]{\includegraphics[width=.45\linewidth]{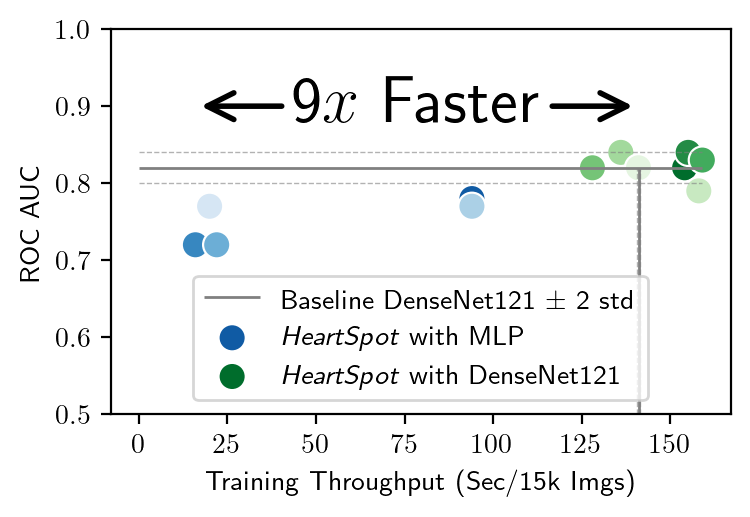}}\\
  \subfloat[Privatized and Explainable\label{fig:1c}]{\includegraphics[width=3.7cm,height=3cm]{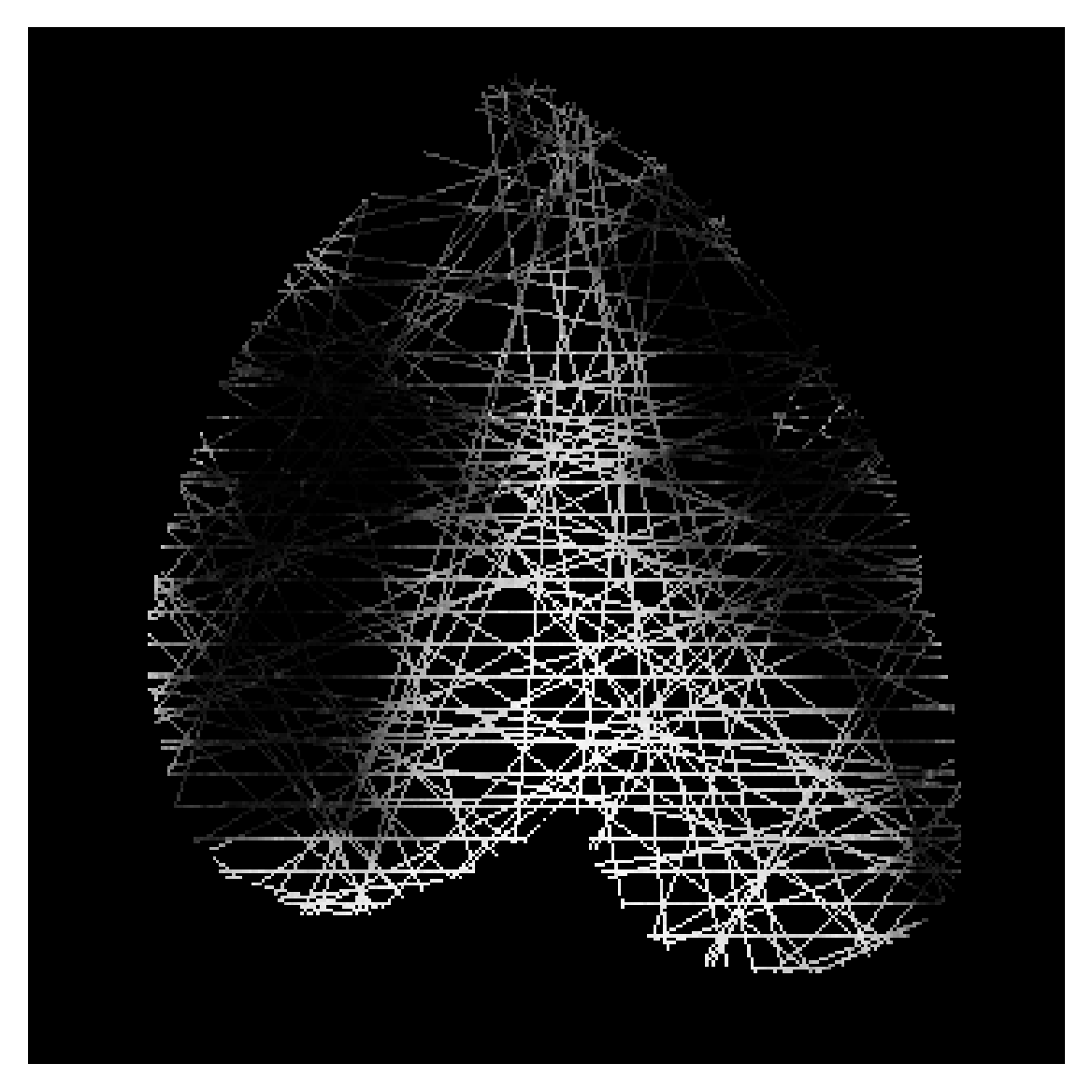}}
  \quad
  \subfloat[Post-hoc Explainable\label{fig:1d}]{\includegraphics[width=3.7cm,height=3cm]{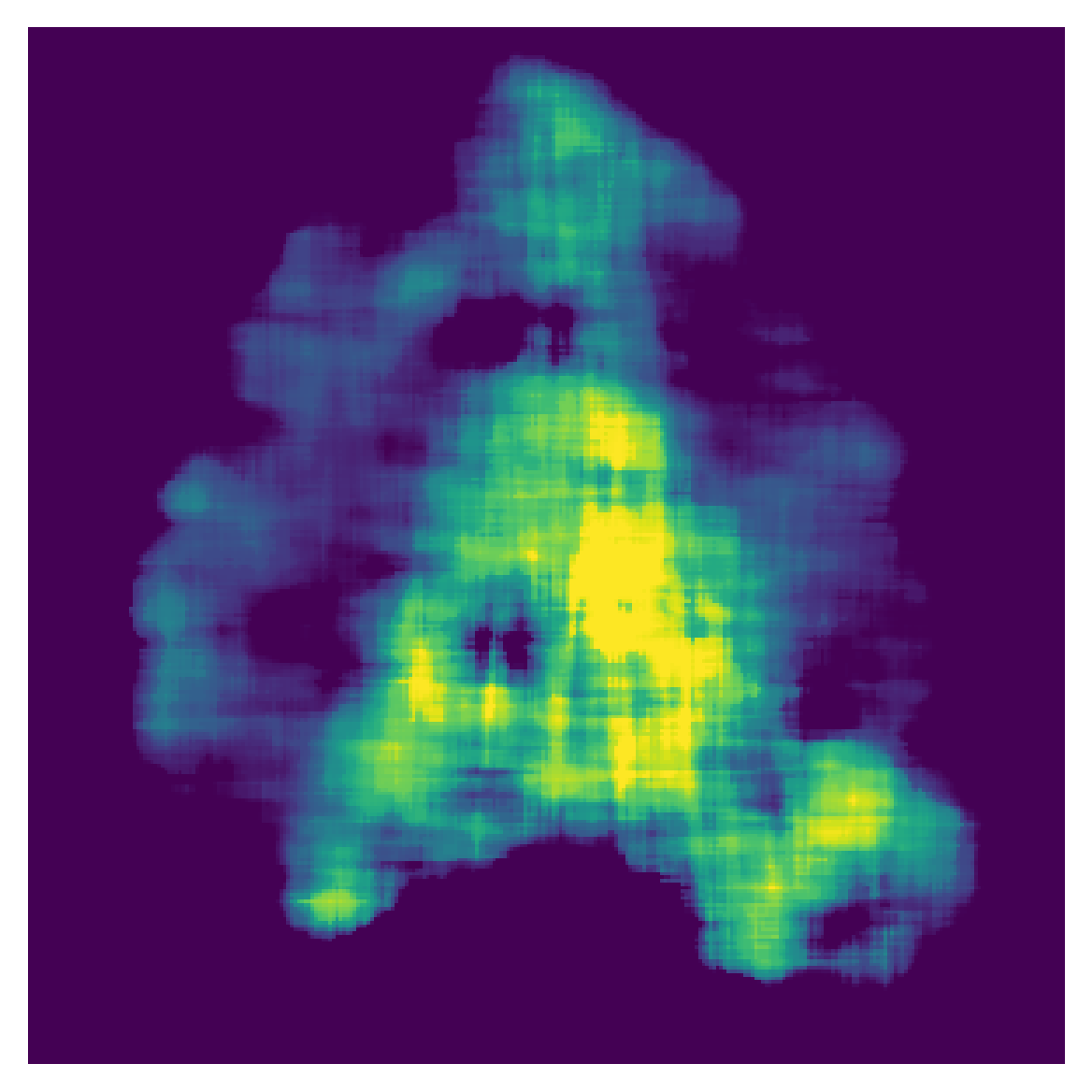}}
  \caption{\method compresses with no learning (\ref{fig:1b}) to give a fast and accurate (\ref{fig:1a}) model.  \method image compression is ante-hoc explainable and privatized (\ref{fig:1c}) with useful post-hoc saliency map explanations obtained without access to the original chest X-ray image (\ref{fig:1d}).}
  \label{fig:fig1perf}
\end{figure}
\begin{figure*}[t]
  \centering
  \includegraphics[width=\textwidth]{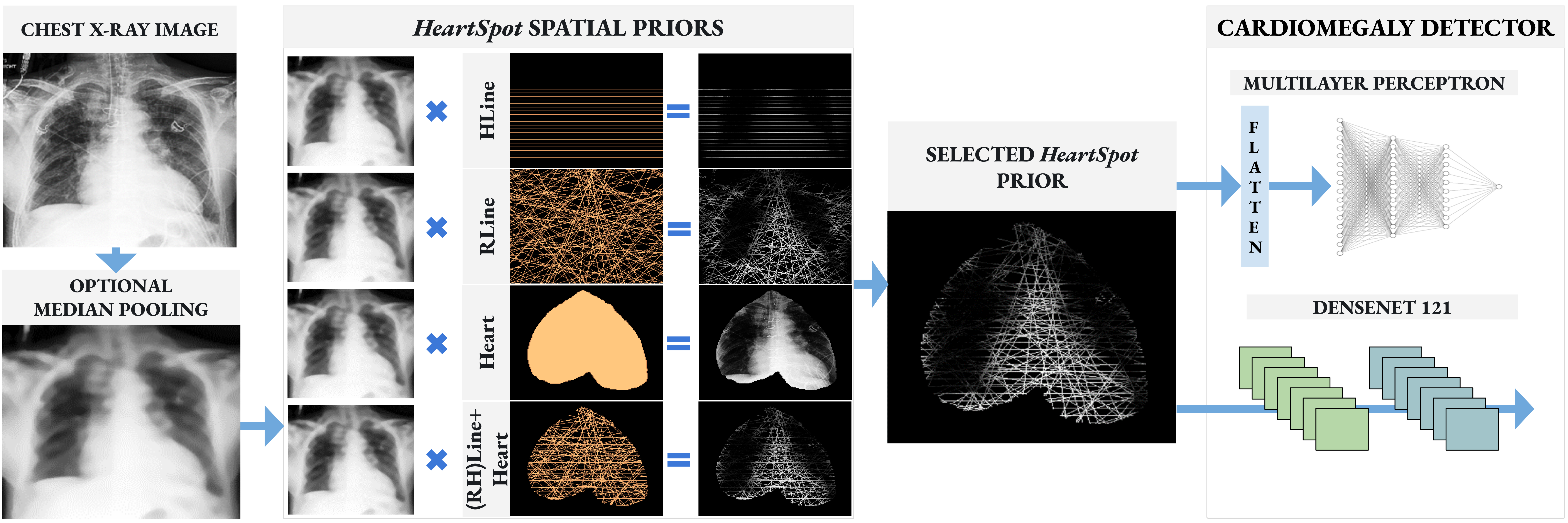}
  \caption{\textbf{\method Cardiomegaly detection pipeline} employs a spatial bias prior to perform pixel sampling.}
  \label{fig:mainfig}
\end{figure*}
As in Figure \ref{fig:fig1perf}, the main contributions of \method are:
\contributions


\section{Methods}
\method performs privatized and explainable image compression for Cardiomegaly detection, as described in Figure \ref{fig:mainfig}.  Domain knowledge is encoded by four spatial bias priors that define pixel sampling strategies.  The sampled pixels give a compressed image or vector for Cardiomegaly detection.

\textbf{Horizontal Lines (HLine).}
The cardio-thoracic ratio compares the transverse diameter of the heart to that of the thoracic cage.  The transverse diameter is the maximum horizontal length between any two heart pixels.  In a 2-d X-ray image, only four points are necessary: two points define a transverse heart diameter, and two describe the thoracic cage width.  Lacking knowledge of where these points are, we sample horizontal lines, or rows of the X-ray image, as shown in the top row of Figure \ref{fig:mainfig}.  We skip the top and bottom of the image based on prior knowledge that the heart does not appear in those regions.   In $320 \times 320$ X-ray images, considering rows 100 to 300 in increments of ten reduces the 102400 pixel image to a 6400 pixel flat vector, giving a $16x$ compression.      
 
\textbf{Random Lines (RLine).}
An alternative cardio-thoracic ratio compares the area of the heart to the area of the lungs.  Assuming the lungs and heart have a smooth perimeter, randomly sampling lines of pixels at arbitrary orientations largely preserves area information.  Lines are randomly sampled by choosing pairs of points uniformly on a circle, with the constraint that each pair contains one point on the left half-circle and the other on the right half-circle.  Uniform sampling of a 2-d euclidean point $[x,y]$ on the unit circle is obtained via $[x,y] = \frac{\mathbf{r}}{||\mathbf{r}||_2}$ where $\mathbf{r} = [r_x, r_y] \sim \mathcal{N}(0,1)$. The point can be forced on the left or right half-circle by updating $x \gets |x|$ or $x\gets -|x|$.  We ensure the lines extend outside the unit circle by multiplying by the image side length. Finally, given pairs of points, the lines are drawn via the Bresenham algorithm.  The random lines are chosen via a random seed, and all images are sampled the same way, as shown in the second row of Figure \ref{fig:mainfig}.  In a $320\times320$ image, selecting only the pixels that intersect with 200 lines gives a flat vector of $3.6x$ fewer pixels.

  \begin{figure}[b]
    \centering
    \includegraphics[width=.368\linewidth,trim=0 -2 0 2]{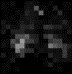}
    \quad
    \includegraphics[width=.4\linewidth]{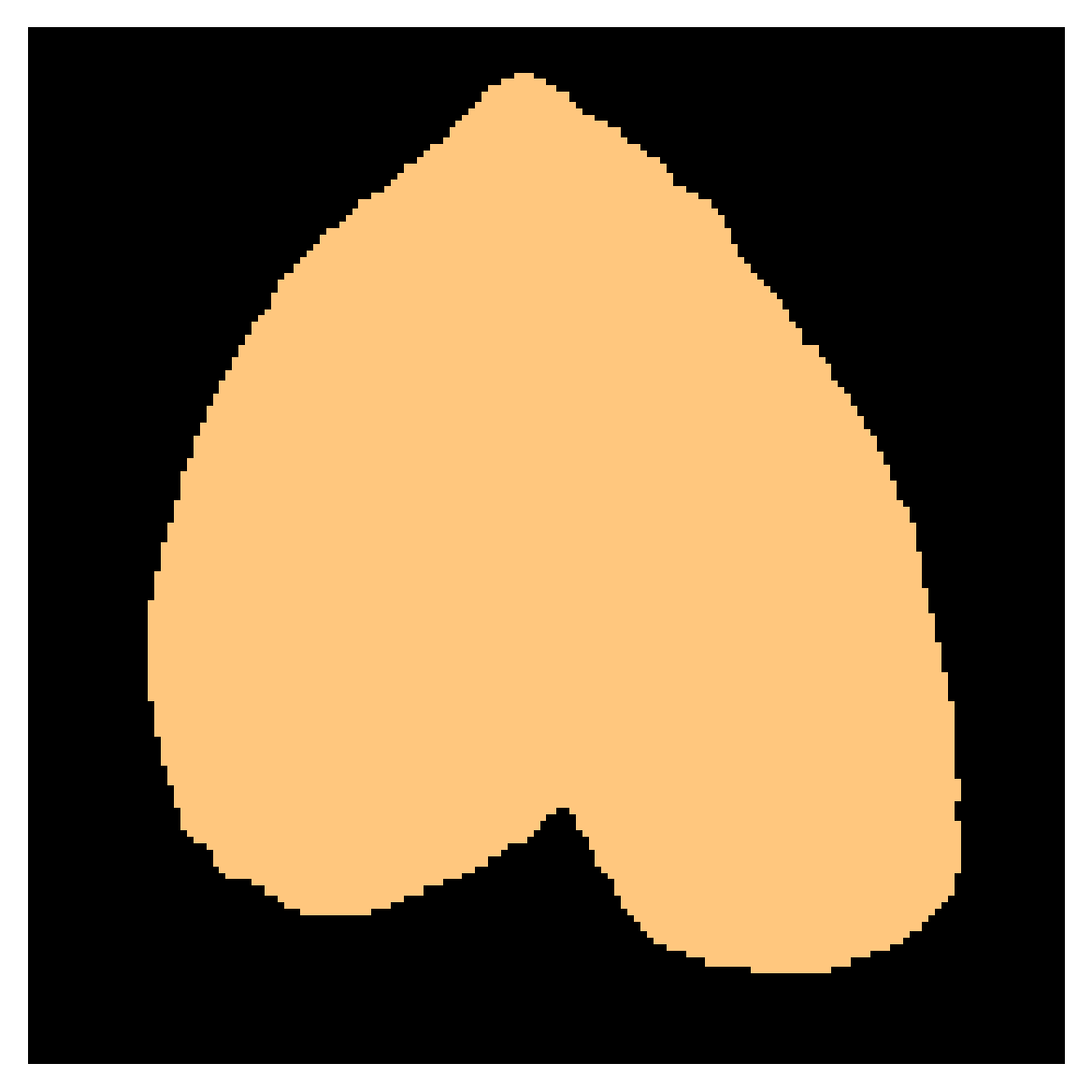}
    \caption{Learned domain knowledge from deep network saliency attributions averaged over multiple X-ray images.}
    \label{fig:saliency}
  \end{figure}

\begin{table*}[t]
\begin{minipage}[t]{.68\linewidth}
\caption{\method Compression Improves Speed and Accuracy}
\label{table:auc_results}
\centering
\begin{tabular}{cccccc}
 \toprule
 & \method & &\multicolumn{2}{c}{Predictive Perf} & Training Speed \\
   & Method & Classifier & ROC AUC & Balanced Acc  & (sec/15k imgs)\\ [0.5ex] 
 \midrule
 \multirow{5}{*}{\rotatebox[origin=c]{90}{\textbf{\;\;\;Fast}}}
   & HLine & MLP &                        0.722 &            0.624    & \textbf{16} \\ 
   & RLine & MLP &             \underline{0.772}&    \textbf{0.703}   & 20  \\  
   & Heart & MLP &                        0.722 &            0.652    & 22  \\ 
   & (RH)Line+Heart &MLP&         \textbf{0.781}&\underline{0.668}    & \underline{17}  \\ 
   & MP+(RH)Line+Heart&MLP&\underline{0.771}&\underline{0.666}   & 94\\  
 \midrule
 \multirow{5}{*}{\rotatebox[origin=c]{90}{\textbf{Accurate}}}
   & HLine & DenseNet121 &                \textbf{0.842} &   \textbf{0.714} & 155  \\
   & RLine & DenseNet121 &                         0.794 &          {0.699} & 158  \\ 
   & Heart & DenseNet121 &                         0.829 &          {0.699} & 159 \\ 
   & (RH)Line+Heart &DenseNet121&                  0.819 &\underline{0.706} & 155 \\ 
   & MP+(RH)Line+Heart &DenseNet121&           0.816 &   \textbf{0.714} & \textbf{128} \\ 
   & MP+HLine &DenseNet121&         \underline{0.837}&\underline{0.706} & \underline{136} \\ 
 \midrule
 \midrule
   & \multicolumn{2}{c}{Baseline DenseNet121 without \method}  & 0.829$\pm$0.017 & 0.701$\pm$0.018  & 143 $\pm$ 4 \\  
 \bottomrule
\end{tabular}
\end{minipage}\hfill
\begin{minipage}[t]{.32\linewidth}
  \caption{Saves Space and Memory}
  \label{tab:compression}
  \centering
  \begin{tabular}{ccc}
    \toprule
    \method Method & IMR & ODR\\
    \midrule
     HLine                 &\underline{6\%} &\underline{15}\%\\
     RLine                 &  28\%          & 59\%           \\
     Heart                 &  47\%          & 60\%*          \\
     (RH)Line+Heart        &  17\%          & 38\%           \\
     MP+(RH)Line+Heart &          {7\%} &          {18\%}\\
     MP+HLine &            \textbf{3\%} &    \textbf{9\%}\\
     \bottomrule
     \\
   \end{tabular}
   \flushleft \scriptsize \vspace{-.5cm}{\qquad* Saved as JPEG image rather than LZMA vector.}
   \begin{center}  
\begin{annotatedFigure}
	{\includegraphics[width=0.32\linewidth]{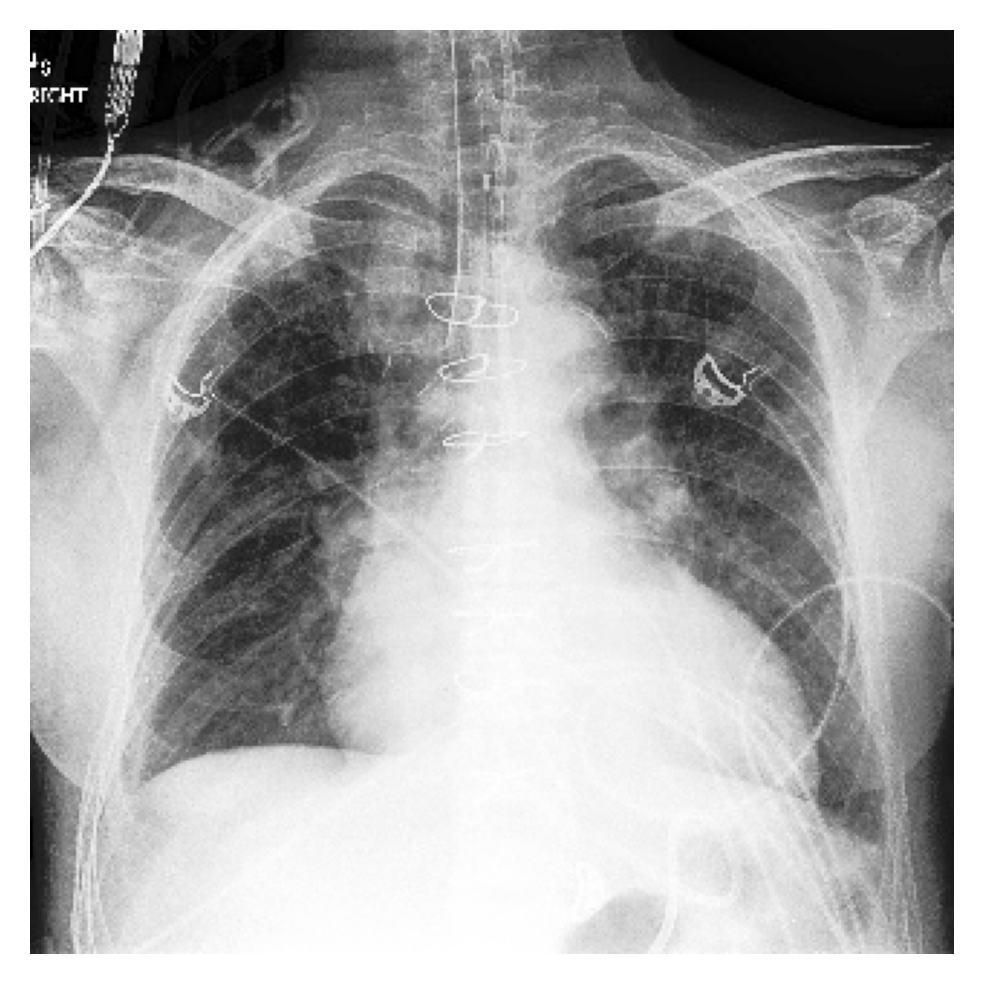}}
	 \pinkFigureBox{0.682,0.5641}{0.8021,0.664}{A}{0.682,0.5641}
	 \pinkFigureBox{0.03,0.756}{0.194,0.972}{B}{0.03,0.756}
	 \pinkFigureBox{0.406,0.496}{0.5352,0.744}{C}{0.406,0.496}
\end{annotatedFigure}
   \includegraphics[width=0.32\linewidth]{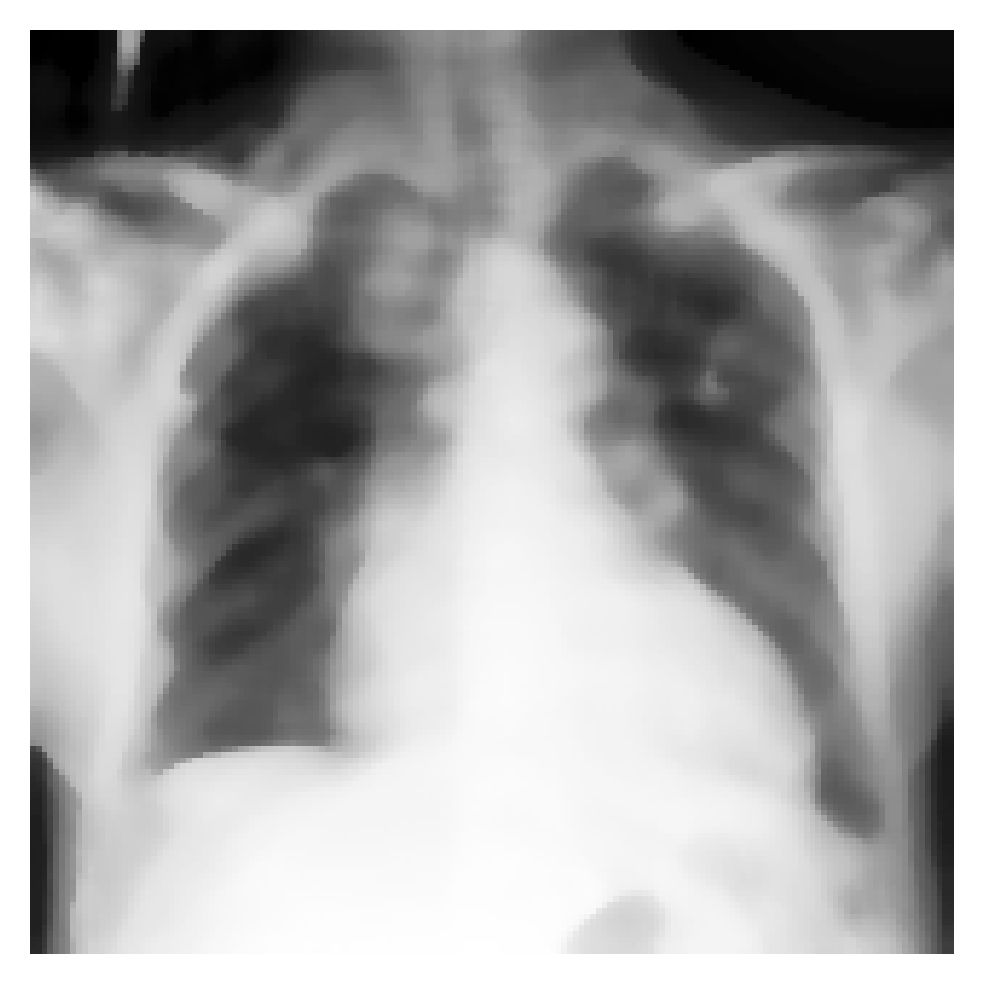}
   \captionof{figure}{\method MP improves privacy.}
   \label{fig:medianpool}
   \end{center}
 \end{minipage}
\end{table*}

\textbf{Region of Interest Heart Mask (Heart).} 
Domain knowledge learned by machines can be obtained by analyzing deep networks with post-hoc saliency attribution methods.  Given a deep convolutional network trained on CheXpert, we average the DeepLift \cite{DeepLift} attributions from multiple validation set X-ray images.  The resulting smoothed explanation describes which pixels the model finds most useful for Cardiomegaly detection.
Figure \ref{fig:saliency} shows the smoothed image and our interpretation of the image as a binary heart mask.  The heart removes all background pixels from the image, giving $2x$ compression.

\textbf{Combined Lines and Heart Masks.}
Combination of all three binary mask images creates a hybrid compression scheme.  The horizontal and random lines employ medical domain knowledge to enable estimation of two cardio-thoracic ratios.  The heart mask represents domain knowledge learned by a neural network.  The lines masks are combined by a set union.  Intersection with the heart mask removes background pixels.  The heart mask improves compression ratios of the lines methods, giving a flat vector of $6x$ fewer pixels.

\textbf{Median Pooling Pre-processing (MP).}
The $k,s$-median pooling filter privatizes images via an edge-preserving smoothing.  MP computes a median of each $k\times k$ patch in the image, as shown in Figure \ref{fig:medianpool} with $k=12$ and stride $s=2$.  In deep networks, the median function is not explicitly modeled in convolution or linear layers, and its subgradient zeros out gradients for all values except one (or two) values in the $k\times k$ kernel.  Our fixed weight compression encoder design has no learning and does not use gradients, making it well suited for a median filter.  For Cardiomegaly detection, edges identifying the heart are of primary interest.  Edge preserving smoothing of MP improves privacy by emphasizing the edges while suppressing other details.  To mitigate the computational cost of median pooling, we use a stride of $s=2$.  The stride downsamples the image by half.

\textbf{Saliency Attribution Maps without Access to the Original Image.}
A saliency attribution of an input tensor gives a flattened saliency tensor of same shape.  With knowledge of the random seed and parameters used to create the combined lines and heart mask, a flattened \method image or corresponding attribution vector can be reconstructed back into an image (e.g. with pseudo-code $\texttt{im = zeros(H,W)};$ $\texttt{im[mask] = flattened\_vector}$).  An example attribution image is shown in Figure \ref{fig:saliency2}.  To facilitate interpretation by a human, we post-process the image by quantile pooling with a large kernel size ($k=24$), 90\% quantile and stride $s=1$.  Quantile pooling generalizes median pooling.  The 90\% quantile imposes a bias to replace sparse values with non-sparse values, and clamp outliers of large value to a more reasonable value.  \method is therefore compatible with standard saliency attribution methods, and does not require access to the original input image.

\section{Results}

\begin{figure*}[t]
  \centering
  \subfloat[Cardiomegaly Positive]{\includegraphics[width=.49\linewidth]{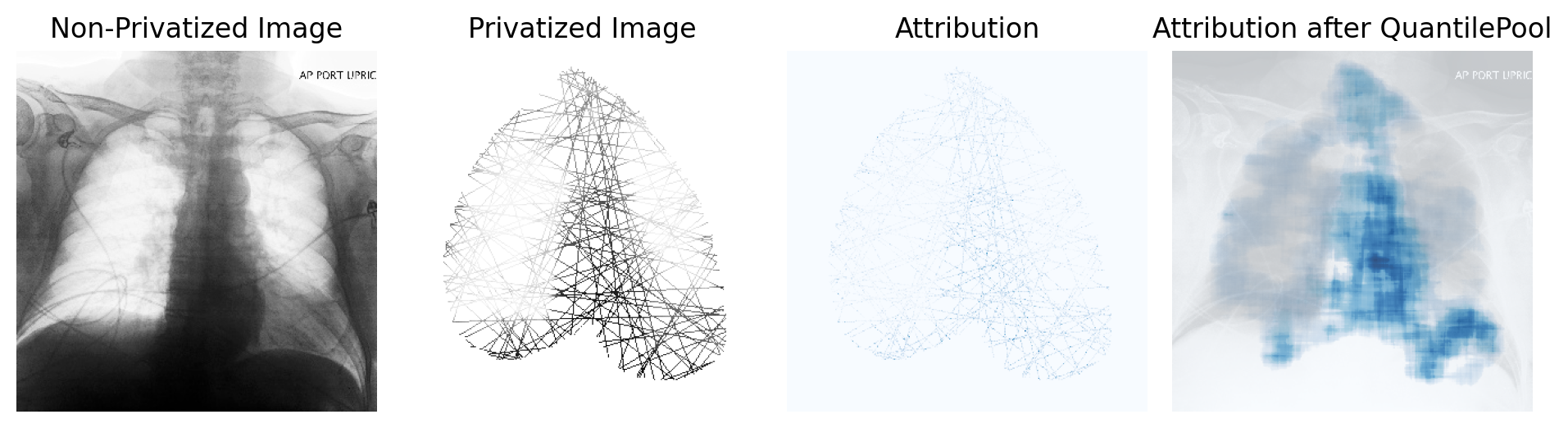}}
  \hfill
  \subfloat[Cardiomegaly Negative]{\includegraphics[width=.49\linewidth]{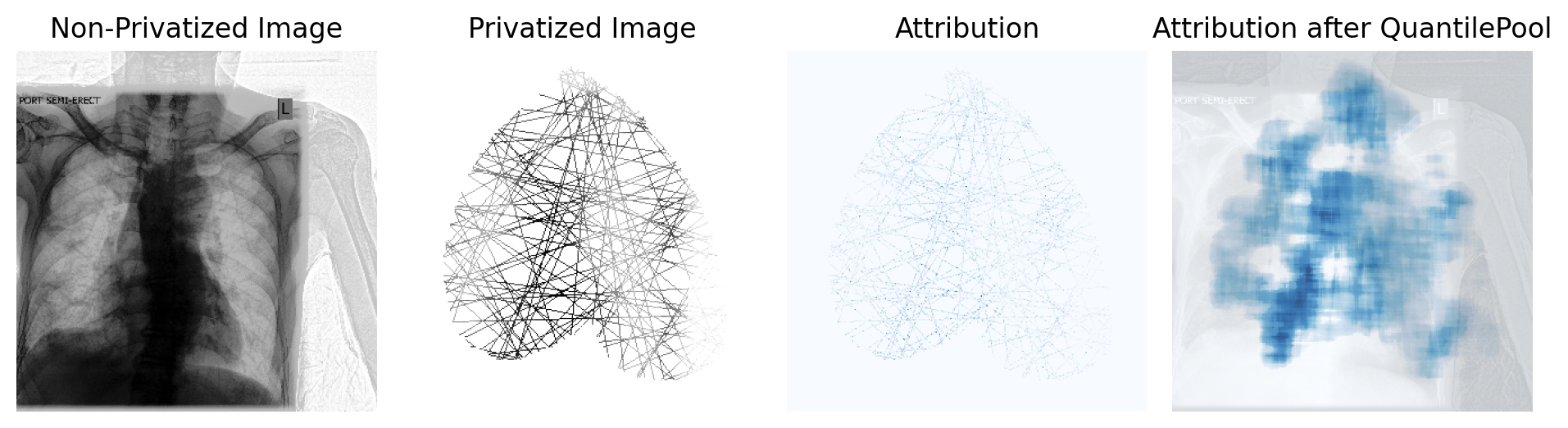}}
  \caption{\textbf{Post-Hoc Saliency Attribution Explanations} without access to the original image.}
  \label{fig:saliency2}
\end{figure*}
Our experiments verify our claims that \method results in fast and performant models, compressed and privatized images, and explanations for Cardiomegaly detection.

\textbf{The CheXpert Dataset} \cite{irvin2019chexpert} contains 220k:234 chest X-ray images from 65k patients in frontal and lateral view.
The 234 image test set has 68 (29\%) Cardiomegaly positive and 166 (71\%) negative samples. The 220k image test set has 13\% positive and 87\% negative samples after ignoring 8k images labeled as uncertain.
We implement the single-model U-Ignore baseline and report the performance on the test set.  The training set was randomly split into 90\% train and 10\% validation, where all images of any given patient fall into only one split.  We include lateral view images.  We use the 11GB CheXpert dataset and center crop all images to $320\times320$.  Each training iteration randomly samples 15k images without replacement.

\textbf{Hyperparameters.} All models were trained with the Adam optimizer, a learning rate of 0.001, and unweighted binary cross entropy loss.  We consider two classifiers: DenseNet121  \cite{densenet} expects an image as input, and a two hidden layer neural network (MLP) expects a flattened vector.  DenseNet121 is commonly used by state-of-the-art works analyzing CheXpert.  The MLP architecture has linear layers with 3200 and 300 hidden nodes respectively, and following each layer is BatchNorm and a SELU activation.  The MLP and DenseNet models were trained for 300 iterations, and the iteration with highest validation area under the ROC curve (ROC AUC) was used for evaluation on the 234 image test set.  Models that use Median Pooling (MP) have a stride of two and kernel size of twelve.  Random Lines (RLine) models have 200 random lines.  Horizontal Lines (HLine) are between 100 to 300 pixels with a 10 pixel step; when Median Pooling with stride two is applied, the range and step is 50:150:5.  Source code is available \cite{our_source_code_github}.

\textbf{Fast and Accurate.}
Table \ref{table:auc_results} and Figure \ref{fig:fig1perf} show that \method 
improves performance of baseline DenseNet121 by up to 0.01 ROC AUC, and that \method with a two layer MLP processes images faster (by up to $9x$).
Table \ref{table:auc_results} reports the test set area under the ROC curve (ROC AUC) and Balanced Accuracy (BAcc), as well as the seconds spent to train 15k images.  The threshold for BAcc was chosen from a validation set ROC curve as the threshold maximizing the true positive rate minus false positive rate.  The Baseline DenseNet121 was trained seven times independently, and we report the test set average and one standard deviation.
For the MLP, we observe that \method (RH)Line+Heart is preferred, giving 17 seconds per epoch ($8x$ faster than DenseNet121) and .78 AUC ROC.
With DenseNet121, the \method MP+HLine method performs best at .84 ROC AUC, a $+0.01$ improvement over the baseline.
\method improves speed and accuracy.

\textbf{Smaller Images.} Table \ref{tab:compression} reports the in-memory compression ratio (IMR) and on-disk ratio (ODR).  A small ratio is better.
  IMR is defined as the number of pixels in the compressed representation divided by the number of pixels in the original center-cropped image.  ODR compares the filesize of a \method flattened vector saved in LZMA format (with default compression preset six) to the filesize of the original center-cropped image saved as JPEG (quality 95).  With knowledge of the \method hyperparameters, the flattened and image representations are interchangeable, offering more flexibility in choice of on-disk compression method.
Note that Heart was saved as JPEG, as LZMA gave 94\% ODR.  MP+(RH)Line+Heart discards 93\% of pixels and is 18\% of the original JPEG filesize when saved on disk, and MP+HLine discards 97\% of pixels and is 9\% of original JPEG filesize.
\method saves memory and disk space via single-image fixed weight compression.

\textbf{Privacy and Ante-Hoc Explainability.}
The \method image in Figure \ref{fig:1c} visually demonstrates privacy preservation by pixel sampling.  All four spatial priors remove identifying features like the collar bone, ribs and shape of the lungs, and high or medium frequency detail is also difficult or impossible to see.  The Median Pooling pre-processing further effectively removes many details, including small lesions, support devices, identifying textures, or shapes of the vertebrae.  By design, \method visually privatizes images yet also remains ante-hoc explainable.  The outline of the heart is still clearly visible, and the prior determines which regions and pixels of the image are relevant.

\textbf{Post-hoc Attribution.}
Figure \ref{fig:saliency2} shows the reconstruction of a saliency attribution vector using the IntegratedGradients method with SmoothGrad, commonly available in the PyTorch Captum library \cite{kokhlikyan2020captum}.  Quantile pooling improves the interpretability of the reconstructed saliency image.  The attribution shows an emphasis on the boundary of the heart and verifies that regions outside the heart were not considered.

\section{Conclusion} 
For privatized compression of chest X-ray images for Cardiomegaly detection, we propose \method with four spatial bias priors.  We recommend \method MP+HLine for best accuracy, compression and privacy.  \method contributions:
\contributions
In future work, \method extends naturally to consideration of other pathologies such as Pleural Effusion and Consolidation.

\printbibliography

\end{document}